\documentstyle[prb,aps,preprint]{revtex}  
\begin{document}
\tighten
\draft 
\title{Ab initio treatment of electron correlations in polymers: 
lithium hydride chain and beryllium hydride polymer}
\author{Ayjamal Abdurahman,$^{1}$~\cite{email} 
Alok Shukla,$^{2}$~\cite{add1} and Michael Dolg$^{1}$\cite{add2}}
\address{$^1$ Max-Planck-Institut f\"ur Physik komplexer Systeme,
N\"othnitzer Str. 38, D-01187 Dresden, Germany}
\address{$^2$Department of Physics and The Optical Sciences Center, 
University of Arizona, Tucson, AZ 85721}
\maketitle



\begin{abstract}
Correlated {\em ab initio\/} electronic structure calculations are 
reported for the polymers lithium hydride chain $[LiH]_{\infty}$ and 
beryllium hydride $[Be_{2}H_{4}]_{\infty}$. First, employing 
a Wannier-function-based approach, the systems are studied
at the Hartree-Fock level, by considering chains, simulating the infinite 
polymers. Subsequently, for the model system $[LiH]_{\infty}$, 
the correlation effects are computed by considering virtual 
excitations from the occupied Hartree-Fock Wannier functions 
of the infinite chain into the complementary space of localized unoccupied 
orbitals, employing a full-configuration-interaction scheme.
For $[Be_{2}H_{4}]_{\infty}$, however, the electron correlation
contributions to its ground state energy are calculated by considering finite 
clusters of increasing size modelling the system. Methods such as
M$\o$ller--Plesset second--order perturbation theory and coupled--cluster singles, 
doubles and triples level of theory were employed. Equilibrium geometry, cohesive 
energy and polymerization energy are presented for both polymers, and 
the rapid convergence of electron correlation effects, when based upon a 
localized orbital scheme, is demonstrated.
\end{abstract}


\section{Introduction}
\label{intro}
Polymers represent a class of one--dimensional infinite crystalline systems 
where {\em ab initio\/} Hartree--Fock (HF) self--consistent field (SCF) 
methods are well developed~\cite{ladik}. An available program 
package is CRYSTAL~\cite{crystal}. However, in order to be able to 
calculate the structural and electronic properties of polymers with 
an accuracy that allows a meaningful comparison with experiment, 
it is usually necessary to include the effects of electron correlations 
into the theory. The most widely used approach here is density--functional 
theory (DFT). Despite its indisputable success in solid state physics 
and computational chemistry as a computationally cheap routine tool 
for large-scale investigations, DFT has the drawback that results depend 
highly on the chosen functional, and cannot be improved in a systematic way.
Wave-function--based quantum chemical {\em ab initio } techniques on the
other hand are free from this flaw, and provide a large array of 
methods of different accuracy and computational cost. Thus it is 
desirable to extend their applicability to infinite systems such as polymers.

Electron correlations are mostly a local effect and therefore 
localized molecular orbitals are preferable to the canonical HF solutions 
for the treatment of large molecules.~\cite{hampel} Similarly, in infinite
systems (localized) Wannier functions provide a better starting point 
for an {\em ab initio\/} treatment of electronic correlations than the
(canonical) Bloch functions. Previous studies of polymers obtained the 
Wannier orbitals from an {\em a posteriori\/} localization of the Bloch 
functions according to a given prescription.~\cite{ladik}
During the last years, in our group a HF approach was developed 
which allows the direct determination of Wannier orbitals within the 
SCF process.~\cite{shukla1}
Various applications to one- and three-dimensional infinite systems 
proved the numerical equivalence of our Wannier--function--based HF approach 
to the conventional Bloch--function--based 
counterpart.~\cite{shukla2,shukla3}

In this paper HF--SCF calculations and subsequent correlation energy calculations 
are presented for the lithium hydride chain $[LiH]_{\infty}$ and the beryllium 
hydride polymer $[Be_{2}H_{4}]_{\infty}$. As a simple, but due to its ionic 
character, nontrivial model polymer, the lithium hydride chain system has 
been previously dealt with in a number of studies.~\cite{shukla3,teramae,tunega}. 
In the present contribution we extend our previous calculation ~\cite{shukla3} to
a wave-function-based {\em ab initio\/} study of electron correlation effects
using a combination of the full configuration interaction (FCI) method and the
the so-called incremental scheme.~\cite{stoll1,stoll2,stoll3} 
The latter approach consists basically in an expansion of the total correlation 
energy per unit cell in terms of interactions of increasing complexity among the 
electrons assigned to localized orbitals (Wannier functions) comprising the 
polymer under consideration.
The electron correlation energy increments needed to establish the total
energy per unit cell are evaluated by considering virtual excitations from a 
small region of space in and around the reference cell, keeping the 
electrons of the rest of the crystal frozen at the Hartree--Fock (HF) level. 
The fast convergence of the incremental expansion allows to truncate it at
relatively low order and thus to calculate the correlation energy of an
infinite system without modelling it as a finite cluster.
However, neither the FCI method nor the incremental approach based on
polymer Wannier orbitals can at present be used for systems with a more
complicated unit cell. Therefore, the second system investigated by us, the  
beryllium hydride polymer, was treated at the coupled-cluster (CC) and
M$\o$ller--Plesset second--order perturbation (MP2) level of theory. Starting 
from the Wannier HF data the correlation corrections to the total energy 
per unit cell were derived from quantum chemical calculations of 
finite model systems using the MOLPRO molecular orbital 
{\em ab initio\/} program package.~\cite{molpro} 
To our knowledge, this system was studied at the HF level two decades ago
by Karpfen using the crystal orbital method, i.e., without including
correlation effects.~\cite{karpfen} Recently, its monomer beryllium
dihydride $BeH_{2}$ has been  well characterized theoretically using reliable 
ab initio and density functional theory methods.~\cite{hinze,jursic} 

The remainder of the paper is organized as follows. In section \ref{methods} 
the applied methods are briefly described. The calculations and results are 
then presented in section \ref{results}. Finally, a summary is given in 
section \ref{summary}. 

\section{Applied  methods}
\label{methods}
Section \ref{wanhf} gives brief outline for the theory within a 
restricted HF (RHF) framework. Sections \ref{inc}
and \ref{sa}, respectively, describe the incremental 
scheme and a simple approach, to compute electron correlation effects in 
polymers.  
\subsection{Wannier--orbital--based Hartree--Fock approach}
\label{wanhf}
Our approach, described in more detail in previous publications.~\cite
{shukla1,shukla2,shukla3} is based upon the direct determination
of the orthonormal Wannier--type (localized) 
orbitals for the polymer. Denoting by $\mid\alpha(\mathbf{R}_{j})\rangle$ 
the Wannier orbitals of a unit cell located at lattice 
vector $\mathbf{R}_{j}$, the set $\{ |\alpha({\bf R}_{i})\rangle;
{\alpha} = 1, n_c; j = 1, N \}$ 
spans the occupied HF space. 
 Here, $n_{c}$ is the number of orbitals per unit cell, 
and $N (\to \infty)$ is the total number of unit cells in the system.
In our previous work we showed that one
can obtain $n_c$  RHF Wannier  functions, $\{|\alpha \rangle, \; \alpha =1,n_c\}$ occupied by $2n_c$ electrons 
localized in the reference unit cell (denoted ${\cal C}$) by solving the 
equations~\cite{shukla1,shukla2,shukla3} 
\begin{equation}
( T + U
 +   \sum_{\beta} (2 J_{\beta}-  K_{\beta})   
+\sum_{k \in{\cal N}} \sum_{\gamma} \lambda_{\gamma}^{k} 
|\gamma({\bf R}_{k})\rangle
\langle\gamma({\bf R}_{k})| ) |\alpha\rangle 
 =  \epsilon_{\alpha} |\alpha\rangle
\mbox{,}
\label{eq-rhf}         
\end{equation}  
where $T$ represents the kinetic-energy operator, $U$ represents
the interaction of the electrons of ${\cal C}$  with the nuclei
of the whole of the crystal, while $J_{\beta}$, $K_{\beta}$,  
respectively, represent the Coulomb and exchange interactions felt
by the electrons occupying the $\beta$-th Wannier function   
of ${\cal C}$, due to the rest of the electrons of the infinite system.
The first three terms of Eq.(\ref{eq-rhf}) constitute the canonical 
Hartree-Fock operator, while the last term is a projection
operator which makes the orbitals localized in ${\cal C}$ orthogonal to those 
localized in the unit cells in the immediate neighborhood of ${\cal C}$
by means of infinitely high shift parameters $\lambda_{\gamma}^{k}$'s. These
neighborhood unit cells, whose origins are labeled by lattice vectors
${\bf R}_{k}$, are collectively referred to as  ${\cal N}$. The 
projection operators along with the shift
parameters play the role of a localizing potential in the Fock matrix, and 
once self-consistency has been achieved, the occupied eigenvectors of 
Eq.(\ref{eq-rhf})  are localized in ${\cal C}$, and are orthogonal to the 
orbitals of ${\cal N}$---thus making them Wannier 
functions~\cite{shukla1,shukla2,shukla3}. As far as the
orthogonality of the orbitals of ${\cal C}$ to those contained in unit cells
beyond ${\cal N}$ is concerned, it should be automatic for systems with
a band gap once  ${\cal N}$ has been chosen to be large enough. Based upon
our past experience regarding a suitable
choice of ${\cal N}$,~\cite{shukla1,shukla2,shukla3} in the 
present calculation we
included up to the third nearest-neighbor unit cells in ${\cal N}$.
For the details concerning the computation
of various terms involving lattice sums ($U$, $J$, and $K$) involved in
Eq. (\ref{eq-rhf}) for the
case of polymers, we refer the reader to  
reference~{\cite{shukla3}}.

\subsection{Incremental method}
\label{inc}
Electron correlation effects in the ground states of a large number of 
three-dimensional ionic and covalent solids,~\cite{inc-cal} as well as 
polymers~\cite{yu} have been studied with the incremental scheme. 
All these calculations used localized orbitals of finite clusters as a
basis set for the correlation treatment. In the present work on the 
lithium hydride chain we use directly the Wannier--functions of the infinite
system. A related study of the three--dimensional lithium hydride solid 
has been published elsewhere~\cite{lih}.

The correlation energy per unit cell is expanded as
\begin{equation}
E_{corr} = \sum_{i} \varepsilon_{i}
+ \sum_{<ij>} \Delta\varepsilon_{ij}
+ \sum_{<ijk>} \Delta\varepsilon_{ijk}
+ ... \label{eq-inc}
\end{equation}
where the summation over $i$ involves Wannier functions located the 
reference cell, while  those over $j$ and $k$ include all the Wannier
functions of the crystal.  
The ``one--body" increments $\varepsilon_{i}$ = $\Delta\varepsilon_{i}$
are computed by considering virtual excitations only from the $i$-th 
Wannier function, freezing the rest of the polymer at the HF level.
The ``two--body" increments $\Delta\varepsilon_{ij}$ are defined as 
$\Delta\varepsilon_{ij}$ = 
$\varepsilon_{ij}-(\Delta\varepsilon_{i}+\Delta\varepsilon_{j})$ 
where $\varepsilon_{ij}$ is the correlation energy of the system obtained 
by correlating two distinct Wannier functions ${i}$ and ${j}$. Thus 
$\Delta\varepsilon_{ij}$ represents the correlation contribution 
of electrons localized on two ``bodies" ${i}$ and ${j}$. 
Higher--order increments are defined in an analogous way. 
Finally, summing up all increments, with the proper
weight factors (according to their occurrence in the unit cell 
of the polymer), one obtains the exact correlation energy per 
unit cell of the infinite system.
In order to get reliable results a size--extensive correlation 
method should be used, although non size--extensive schemes 
also may provide reasonable estimates if the
incremental expansion is truncated at low order.
In the present work for the lithium hydride chain we choose the strictly
size-extensive full configuration interaction (FCI) method. 
As mentioned earlier, when computing the correlation contributions via
Eq. (\ref{eq-inc}), except for the orbitals involved (say orbitals $i$ and $j$
for the two-body increment $\Delta \epsilon_{ij}$), the rest of the occupied 
Wannier orbitals of the
infinite solid are held frozen at the HF level. The region
containing these frozen orbitals plays the role of the ``environment''
for the electrons involved in the correlated calculations, and its
contribution  can be absorbed in the so-called ``environment potential''
$U^{\mbox{env}}$ defined as 
\begin{equation}
 U^{\mbox{env}}_{pq}= \sum_{\alpha({\bf R}_{j}) \in {\cal E} } 
( 2 \langle p \alpha({\bf R}_{j})|\frac{1}{r_{12}}|q \alpha({\bf R}_{j}) \rangle
 - \langle p \alpha({\bf R}_{j})|\frac{1}{r_{12}}|\alpha({\bf R}_{j}) q \rangle
 ) \; \mbox{,} \label{eq-uenv}
\end{equation}
where ${\cal E}$ represents the unit cells of the environment, $p$ and $q$ 
are two arbitrary basis functions, and the factor of
two in the first term is due to the spin summation.
The sum of Eq.(\ref{eq-uenv}) involves infinite lattice sum over the
environment unit cells, and is computed
by simply subtracting from the lattice summed $J$ and $K$ integrals (cf.
Eq. (\ref{eq-rhf})) obtained at the
end of the HF iterations, the contributions corresponding to the orbitals
being correlated. Once 
$U^{\mbox{env}}_{pq}$ has been computed, one is left with an effective
Hamiltonian involving a finite number of electrons located in the
region whose Wannier orbitals are being correlated. Physically speaking  
$U^{\mbox{env}}_{pq}$  represents the influence of the environment electrons 
on the electrons being correlated, explicitly.
In the present calculations the Li $1s^2$ core shell was also kept frozen, and its
contribution was also included in $U^{\mbox{env}}_{pq}$. The basis functions
$p$ and $q$ were restricted to those of the reference cell and the adjacent cells 
up to the third-nearest neighbors.

The virtual orbitals used for computing the correlation effects were also
localized. They were obtained by first orthogonalizing the basis set
to the occupied space by using  corresponding projection operators, as 
suggested by Pulay.~\cite{pulay} Subsequently the basis functions
are orthogonalized to each other using the symmetric-orthogonalization
procedure, yielding a localized and orthonormal virtual orbital set.~\cite{lih} 
The number of virtual orbitals per unit cell considered for a specific
increment corresponds to the number of basis functions per unit cell
minus the number of occupied orbitals per unit cell. The virtual orbitals have
been expanded in the same basis set as described above for $U^{\mbox{env}}_{pq}$.
\subsection{A simple approach}
\label{sa}
In principle the total energy $E_{tot}$ per $[Be_{2}H_{4}]$ unit cell of 
beryllium hydride may be obtained as the limit
\begin{equation}
E=\lim_{n \to \infty}{E(Be_{2n+1}H_{4n+2})\over n} ,
\end{equation} 
i.e., by performing calculations for increasingly long oligomers 
$H(BeH_{2})_{2n}BeH$. In order to reduce finite-size effects due to the 
termination of the oligomers by one beryllium and two hydrogen atoms 
saturating the dangling bonds of ${\cdot } (BeH_{2})_{2n} {\cdot }$ , one 
may consider instead
\begin{equation}
E=\lim_{n \to \infty}\bigtriangleup E_{n}=\lim_{n \to
\infty}\biggl[E(Be_{2n+3}H_{4n+6})-E(Be_{2n+1}H_{4n+2})\biggr],
\label{eq-sa}
\end{equation}
i.e., the energy change between subsequent oligomers differing by 
a single unit cell. Therefore, identical unit cells were used as building
blocks for both oligomers, i.e., the geometrical optimization was 
restricted only to parameters relevant for the polymer beryllium hydride.\\ 
Since the convergence of $\bigtriangleup E_{n}$ with respect to n is much
faster for the correlation contributions than for the HF energy, and HF
programs treating the infinite system are at hand (CRYSTAL, WANNIER), we use
Eq. (\ref{eq-sa}) only for the correlation energy per unit cell. 
This approach has previously been used successfully in calculations 
for trans-polyacetylene,~\cite{yu} and some boron-nitrogen 
polymers.~\cite{ayjamal} 
\section{Calculations and Results}
\label{results}
\subsection{$[LiH]_{\infty}$} 
HF ground state calculations are a necessary prerequisite for the application 
of the incremental approach to electron correlation. 
We performed such calculations for a lithium hydride chain oriented 
along the x-axis using the WANNIER code \cite{wannier}.
The reference cell contained hydrogen at the (0,0,0) and 
lithium at the $({\it a}/2,0,0)$, where ${\it a}$ is lattice constant.
We adopted the extended basis set optimized by Dovesi 
{\em et al.}~\cite{dovesi} 
First, all--electron Wannier HF calculations were performed 
at the different
lattice constants in the range 2.8--4.0 (\AA) and the total HF energy 
per unit cell for various lattice constants 
near the equilibrium was fitted to a cubic polynomial in 
order to derive the ground state HF equilibrium lattice constant 
and total energy.
After determining the Wannier orbitals for each value of the lattice 
constant, the  corresponding 
FCI calculations were performed by means of the incremental scheme. 
The expansion of the correlation energy per unit cell
was restricted to one-- and two--body increments, and included 
interactions up to third--nearest neighbor unit cells. Contributions 
from higher order increments as well as from interactions
between more distant cells proved to be negligible.
The equilibrium values for the FCI energy per unit cell 
and the lattice constant were determined as described for 
the HF results.
The main contribution of 98.8 \% to the correlation energy per unit cell 
at the equilibrium geometry ($E$=$-0.0307 a.u.$) comes from the one-body term.
Two-body terms for first--, second-- and third--nearest neighbors contribute 
with 1.15, 0.01 and 0.001 \%, respectively. Our results are summarized
in table \ref{t1}. It is quite obvious from table \ref{t1} that, as a function
of distance, the two-body correlation effects converge very rapidly.

Since the Li basis set used here is suitable only for the ionic LiH molecule, 
we cannot get 
a good result for the atomic reference energy of the neutral Li atom 
(which is needed to determine the cohesive energy).
 Therefore, for this almost ideally ionic chain the cohesive energies both at the HF and the
correlated level are obtained 
by subtracting the electron affinities (EA) and ionization potential 
(IP) from the dissociation energy calculated with respect to the ions 
$Li^{+}$ and $H^{-}$. The HF values of EA and IP are determined using 
the finite--difference atomic HF program MCHF \cite{mchf}. The experimental 
values of EA and IP were taken as the CI limit, i.e., disregarding the very
small relativistic effects. For the 
polymerization energy we optimized the Li--H distance for the
$^1\Sigma^+$ ground state of the monomer at the HF and CI level.
Our results are summarized in table \ref{t2}. It is clear from table
\ref{t2} that, as expected, correlation effects contribute significantly
to the cohesive energy. However, they do not make any significant contribution
to the lattice constant of the system.
\subsection{$[Be_{2}H_{4}]_{\infty}$}
Beryllium hydride has attracted considerable interest as a rocket fuel on
account of its high heat of combustion. It has also been considered as a
moderator for nuclear reactors. From the previous studies \cite{bery} 
we also know that it is poisonous and difficult to prepare for experiment. 
Even though there is no or very little experimental information about 
the polymer, it has been studied theoretically using reliable 
{\em ab initio\/} methods at the HF level by Karpfen \cite{karpfen}. 
In the present work we have studied this polymer at the HF and the correlated 
level. The Wannier--orbital--based HF--SCF approach,  
coupled-cluster (CC), and M$\o$ller--Plesset 
second--order perturbation (MP2) theory were employed to 
determine the equilibrium structures and total energies per unit cell.
 In our calculations the unit cell included two beryllium and four
hydrogen atoms and has a perfect tetrahedral structure with all four Be-H bond 
distances equal, i.e., there are two HBeHBeH planes that are perpendicular to
each other, with the beryllium atoms in their crossings. 
In the cluster approximation the unit cell is terminated by one beryllium and two
hydrogen atoms. In this structure the terminal beryllium atoms have 
trigonal coordination while all others are distorted tetrahedrons.
 First we optimized the structure 
of this polymer at the HF--SCF level using the CRYSTAL \cite{crystal} program. 
The total HF energies obtained with the CRYSTAL program were then taken as an 
input for a re-optimization at the MP2, CCSD (CC singles and 
doubles) and CCSD(T) (CCSD with a perturbative estimate of triples) level. 
The correlation energy contributions at each geometry have been calculated 
with the MOLPRO  molecular orbital {\em ab initio\/} program 
package \cite{molpro} 
by using the simplified finite--cluster approach in which we 
put {\em n}=3  in Eq. (\ref{eq-sa}). In this system the correlation
energy converges rapidly with respect to cluster size, i.e., for {\em n}=3, 
one finds $\bigtriangleup E_{4}-\bigtriangleup E_{3}{\approx}10^{-6}$ a.u..
We have optimized the  beryllium--hydride bond length $(r_{BeH})$ and 
the lattice constant (a). We adopted polarized valence 
double--zeta ( 6--31G$^{**}$) basis sets for beryllium and for hydrogen. 
The polarization functions 
consisted of a single p--type exponent of $0.75$ Bohr$^{-2}$ on hydrogen and 
single d--type exponents of $0.4$ Bohr$^{-2}$ on beryllium. 
In our HF calculations for polymers we optimized the most diffuse s--type 
exponent, which is less than $0.1$ in the original 6--31G$^{**}$ basis set,
and obtained $0.15$. A smaller value causes linear dependencies in the basis 
set when applied in the infinite system.\\ 
We have also calculated the cohesive energy per unit cell at the HF 
and correlated level. The atomic HF--SCF, MP2, CCSD and CCSD(T) reference 
energies (Be: $-14.5668$ a.u., $-14.5928$ a.u., $-14.6131$ a.u. 
and $-14.6131$ a.u.; H: $-0.4982$ a.u.) were obtained with the original  
6--31G$^{**}$ basis sets. In addition to the cohesive energy, we 
have also calculated the polymerization energy. The geometry of 
the monomers was optimized at 
the SCF, MP2, CCSD, and CCSD(T) 
levels of theory employing the MOLPRO program \cite{molpro}. Our final results are
summarized in table \ref{t3}. Due to the absence of experimental data 
or theoretical results at the correlated level, we compare our 
result only at the HF level. To the best of our 
knowledge, only Karpfen \cite{karpfen} has performed a geometry 
optimization for this polymer within an ab initio crystal 
Hartree--Fock approach and 
his results are also given in table \ref{t3}. Our beryllium--hydrogen 
bond length is in good agreement with the one obtained by Karpfen,
but our HF energy is lower 0.05 a.u. than the value of
Karpfen \cite{karpfen}. A possible reason is the use of d functions
in our basis sets. 
\section{Summary}
\label{summary}
In conclusion, given a well-localized basis set of Wannier orbitals
size-extensive standard quantum chemical methods such as 
full configuration interaction, coupled-cluster or many-body perturbation 
theory can be applied to evaluate ground state properties of polymers.
Rapid convergence of the incremental expansion of the correlation energy
is obtained for ionic systems, e.g., the simple model of the lithium
hydride chain. In beryllium hydride polymer electron correlation 
accounts for 12--14{\%} of the cohesive energy and 22--24{\%} of the 
polymerization energy at all three levels 
of theory and reduces the lattice constant.
In all the cases it was demonstrated that the use of localized orbitals
leads to a rapid convergence of electron correlation effects, thus making it 
possible for one to compute the electron correlation effects of 
infinite systems.

\clearpage
\newpage

\begin{table}
\caption{Various increments to the correlation energy (in
Hartrees) computed by the Wannier-function-based approach
presented in this work. The results refer to
lattice constant of 3.30 (\AA). NN stands for nearest
neighbors.}\label{t1}
\begin{tabular}{lll}
\hline
Correlation & &  Energy \\
Increment   & & \\
\hline
\hline
one-body      & &-0.0303345\\  
two-body (1NN)& &-0.0003538\\   
two-body (2NN)& &-0.0000035\\  
two-body (3NN)& &-0.0000003\\ 
\hline
\hline
\end{tabular}
\end{table}

\begin{table}
\caption{Total energy  E$_{tot}$ (Hartree),
cohesive energy  $\bigtriangleup$E$_{coh}$
(eV), polymerization energy $\bigtriangleup$E$_{pol}$ (eV) per unit cell and
lattice constant a (\AA) of the lithium hydride chain.}
\label{t2}
\begin{tabular}{lllll}
\hline
Method & E$_{tot}$ & $\bigtriangleup$E$_{coh}$ & $\bigtriangleup$E$_{pol}$&a  \\
\hline
\hline
WANNIER SCF&-8.038047&3.8760&1.8067 &3.3273\\  
CRYSTAL SCF&-8.038031&3.8759&1.8063 &3.3274\\   
FCI &-8.068744&4.6545&1.4854 &3.3300\\  
\hline
\hline
\end{tabular}
\end{table}

\begin{table}
\caption{Total energy E$_{tot}$ (Hartree), cohesive energy $\bigtriangleup$E$_{coh}$ (eV), 
polymerization energy $\bigtriangleup$E$_{pol}$ 
(eV) per unit $Be_{2}H_{4}$ and lattice constant a (\AA), Be--H distance h (\AA) of beryllium hydride.}
\label{t3}
\begin{tabular}{llllll}
\hline
Method & E$_{tot}$ & $\bigtriangleup$E$_{coh}$ & $\bigtriangleup$E$_{pol}$&a&h  \\
\hline
\hline
CRYSTAL SCF &-31.6300 &13.70&2.645&3.958&1.467    \\
MP2$^{a}$ &-31.7608&15.85&3.445 &3.958&1.456  \\ 
CCSD$^{a}$ &-31.7908 &15.56&3.402&3.969&1.457     \\    
CCSD(T)$^{a}$ &-31.7944&15.66&3.478&3.968 &1.458  \\
Karpfen$^{b}$&-31.5780&--&--&4.024&1.470\\
\hline
\hline
\end{tabular}

$^{a}$ correlation contributions added to CRYSTAL SCF energies.\\
$^{b}$ performed with $7, 1/4$ basis sets considering third neighbor's 
interactions.\\
\end{table}


\begin{thebibliography}{99}
%
\bibitem[\ast]{email}{email: ayjamal@mpipks-dresden.mpg.de}
\bibitem[\dagger]{add1}{Present address: Department of Physics, Indian 
Institute of Technology, Powai, Mumbai 400 076, India}
\bibitem[\ddagger]{add2}{Permanent address: Institut f\"ur Physikalische und Theoretische
Chemie, Universit\"at Bonn, Wegeler Str. 12, 53115 Bonn, Germany}
\bibitem{ladik} See, e.g., J. J. Ladik, Quantum Theory of Polymers as Solids. 
Plenum press, New York, NY (1988).
%
\bibitem{crystal} R. Dovesi, V.R. Saunders, C. Roetti, M. Causa,
N.M. Harrison, R. Orlando, E. Apra  CRYSTAL95 user's manual. University of
Turin, Italy (1996).
%
\bibitem{hampel} C. Hampel and H. -J. Werner, J. Chem. Phys. {\bf 104}, 6286
(1996), and references cited therein.
%
\bibitem{shukla1}A. Shukla, M. Dolg, H.Stoll and P. Fulde, Chem. Phys. Lett.
{\bf 262}, 213 (1996).
%
\bibitem{shukla2} A. Shukla, M. Dolg, P. Fulde, and H. Stoll, Phys. Rev. B, {\bf 57}, 1471 (1998).
%
\bibitem{shukla3} A. Shukla, M. Dolg, and  H. Stoll, Phys. Rev. B {\bf 58}, 4325 (1998).
%
\bibitem{teramae} H. Teramae, Theor. Chim. Acta {\bf 94}, 311 (1996).
%
\bibitem{tunega} D. Tunega, J. Noga, Theor. Chim. Acta {\bf 100}, 78 (1998).   %
%
\bibitem{stoll1} H. Stoll, Phys. Rev. B {\bf 46}, 6700 (1991).
%
\bibitem{stoll2} H. Stoll, J. Chem. Phys. {\bf 97}, 8449 (1992).
%
\bibitem{stoll3} H. Stoll, Chem. Phys. Lett. {\bf 191}, 548  (1992).
%
\bibitem{molpro}  H.-J. Werner and P. Knowles, MOLPRO, 1994, is a package 
of {\em ab initio\/} programs written by H.-J. Werner and P.J. Knowles, 
with contributions from J.Alml{\"o}f, R. D. Amos, A. Berning, C. Hampel, 
R. Lindh, W. Meyer, A. Nicklass, P. Palmieri, K.A. Peterson, 
R.M. Pitzer, H. Stoll, A.J. Stone, P.R. Taylor.
%
\bibitem{karpfen} A. Karpfen, Theor. Chim. Acta {\bf 50}, 49 (1978).    
%
\bibitem{hinze} J. Hinze, O. Friedrich, and A. Sundermann, Mol. Phys, {\bf 96}, 711 (1999). 
%
\bibitem{jursic} B. S. Jursic J. Mol. Struct. (Theochem) {\bf 467}, 7 (1999).
%
\bibitem{inc-cal}{See, e.g., B. Paulus, P. Fulde and H. Stoll, Phys. Rev. B 
{\bf 54}, 2556 (1996); K. Doll, M. Dolg, P. Fulde, and H. Stoll, Phys. Rev. B
{\bf 55}, 10282 (1997).}
\bibitem{yu} M. Yu, S. Kalvoda, and M. Dolg, Chem, Phys {\bf 224}, 121 (1997).
\bibitem{ayjamal}{A. Abdurahman, M. Albrecht, A. Shukla, and M. Dolg,
J. Chem. Phys. {\bf 110}, 8819 (1999).}
%
\bibitem{lih} A. Shukla, M. Dolg, P. Fulde, and H. Stoll, Phys. Rev. B {\bf
60}, 5211 (1999).
%
\bibitem{pulay}{P. Pulay, Chem. Phys. Lett. {\bf 100}, 151 (1983).}
%
\bibitem{wannier} Computer program WANNIER, A. Shukla, M. Dolg, H. Stoll, and 
P. Fulde (unpublished).
%
\bibitem{dovesi} R. Dovesi, C. Ermondi, E. Ferrero, C. Pisani, and C. Roetti  
Phys. Rev. B {\bf 29}, 3591 (1984).
%
\bibitem{mchf} MCHF atomic electronic structure code, C. Froese-Fischer, 
The Hartree-Fock Method for Atoms -- A Numerical Approach, 
Wiley, New York, 1976. 
%
\bibitem{bery} Ullmann's Encyclopedia of Industrial Chemistry, Fifth,
Completely Revised Edition {\bf A13}, 205 (1989). 
%
\end{thebibliography}
\end{document}